# Generation of rogue waves in gyrotrons operating in the regime of developed turbulence

N. S. Ginzburg,[1] R. M. Rozental,[1] A. S. Sergeev,[1]
A. E. Fedotov,[1] I. V. Zotova,[1*] and V.P.Tarakanov[2]

[1]*Institute of Applied Physics, RAS, Nizhny Novgorod, Russia*

[2]*Moscow Engineering Physics Institute, Moscow, Russia*

*\*zotova@appl.sci-nnov.ru*

**Abstract**

*Within the framework of the average approach and direct 3D PIC (particle-in-cell) simulations we demonstrate that the gyrotrons operating in the regime of developed turbulence can sporadically emit "giant" spikes with intensities a factor of $100 \div 150$ greater than the average radiation power and a factor of $6 \div 9$ exceeding the power of the driving electron beams. Together with the statistical features such as a long-tail probability distribution, this allows the interpretation of generated spikes as microwave rogue waves. The mechanism of spikes formation is related to the simultaneous cyclotron interaction of a gyrating electron beam with forward and backward waves near the waveguide cutoff frequency as well as with the longitudinal deceleration of electrons.*

Many physical systems exhibit behavior associated with the emergence of the so-called rogue waves which represent dramatically high-amplitude events that occur with low probability but much more frequently than expected in ordinary (e.g., Gaussian or Rayleigh) wave statistics. The rogue wave concept has it origins in hydrodynamics [1]. But in recent years, it is generally accepted that rogue waves are ubiquitous in nature [2-9]. Besides water waves, they have been reported in liquid helium [3], nonlinear optics [4], semiconductor and fiber lasers [5-7], plasma systems [8], etc. In microwaves, rogue waves have recently been observed in the study of the electromagnetic wave transport over a plate with randomly placed metal cones [9]. Obviously, it can be suggested that, similar to lasers [5-7], the rogue waves can also occur in microwave oscillators representing active systems which are based on the interaction of electromagnetic waves with non-equilibrium electron beams.

One of the widely studied microwave oscillators is a gyrotron in which a helical electron beam excites a waveguide mode near its cutoff frequency [10]. In this paper, we demonstrate that rogue waves can appear in gyrotrons operating in the regime of developed turbulence. It is well known that in electronic oscillators, including gyrotrons, an increase of the excess over the starting conditions leads to a complication of the radiation spectrum caused by the appearance of periodic self-modulation and then chaotic turbulent generation regimes [11-17]. Typically, realization of self-modulation regimes is a result of combination of a nonlinearity and delay effects [12].

Nevertheless, as we show in this paper, some specific features of the gyroton operation facilitate the unusual chaotic dynamics including the sporadic generation of ultrashort giant pulses that can be interpreted as rogue waves. First of all, we mean the simultaneous cyclotron interaction of an electron beam with forward and backward propagating waves where the gyrofrequency is close to the cutoff frequency (see points *I* and *II* in the dispersion diagram, Fig. 1). One more critical factor which should be included into consideration is the change of the longitudinal momentum of electrons. This effect is not significant for a steady-state generation regime and, as a rule, is ignored in the gyrotron theory [14-18]. However, in the process of the formation of giant ultrashort pulses with a very steep profile, significant gradients of the electric field appear at their fronts. Correspondingly, strong transverse magnetic fields are initiated even near a cutoff frequency, which, in turn, leads to a significant change in the longitudinal momentum of electrons. As a result, the energy of the translational motion transforms to the energy of the transverse rotation of particles and then in the energy of an electromagnetic pulse.

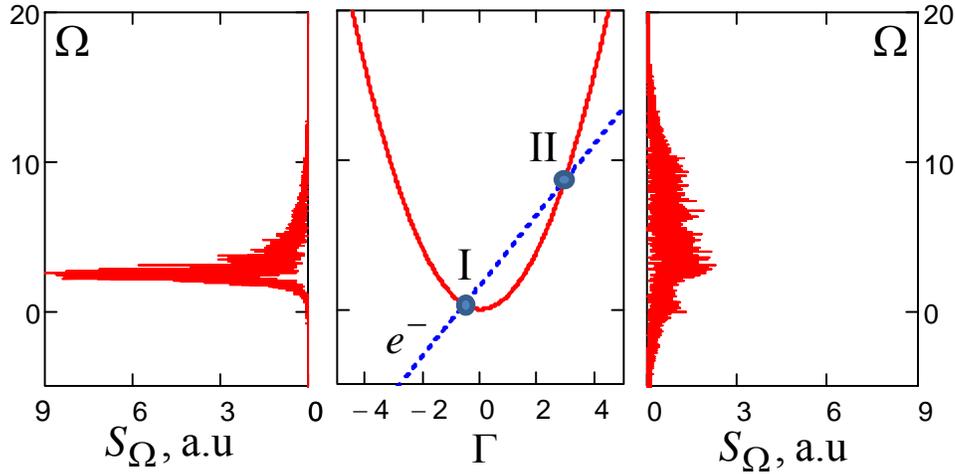

**Fig.1** Dispersion diagram of the waveguide mode $\Omega = \Gamma^2$ and the electron beam line $\Omega = 4g_0^{-2}(\Gamma - \Delta)$ in the regime of rogue wave generation in a gyrotron (center). Radiation spectra at the entrance $Z = 0.2$ (left) and output $Z = 15$ (right) of an interaction space. ($L = 15$, $g_0 = 1.3$, $\beta_{\perp 0} = 0.2$, and $\Delta = -0.7$).

In further consideration we will assume that the interaction space of a gyrotron represents a section of a regular cylindrical waveguide in which a rotating $TE_{mn}$ mode is excited at a quasi-cutoff frequency $\omega_c$ by a beam of electrons gyrating in the homogeneous magnetic field $\vec{H} = H_0 \vec{z}_0$. Correspondingly, the transverse electric and magnetic components of an electromagnetic wave can be represented as

$$\vec{E}_\perp = \kappa^{-1} \text{Re}\left(A(z,t)[\nabla_\perp \Psi \times \vec{z}_0] e^{i\omega_c t}\right), \quad \vec{H}_\perp = \kappa^{-2} \text{Re}\left(i\frac{\partial A(z,t)}{\partial z} \nabla_\perp \Psi e^{i\omega_c t}\right), \qquad (1)$$

where $A(z,t)$ is the slowly varying wave amplitude, $\Psi(r,\varphi) = J_m(\kappa r) e^{-im\varphi}$ is a membrane function, $J_m$ is a Bessel function, and $\kappa = \omega_c/c$. In this case, we can describe the interaction in gyrotrons at the fundamental cyclotron harmonic by a non-uniform parabolic equation for the wave amplitude in combination with averaged electron motion equations for the transverse $p_\perp$ and longitudinal $p_\parallel$ momenta of particles:

$$i\frac{\partial^2 a}{\partial Z^2} + \frac{\partial a}{\partial \tau} = iI_0 J,$$

$$\frac{\partial \hat{p}_\perp}{\partial Z} + \frac{g_0^2}{4}\frac{\partial \hat{p}_\perp}{\partial \tau} + i\frac{\hat{p}_\perp}{\hat{p}_\parallel}\left(\Delta - 1 + |\hat{p}_\perp|^2 + \frac{\hat{p}_\parallel^2 - 1}{g_0^2}\right) = i\frac{a}{\hat{p}_\parallel} + \frac{\beta_{\perp 0}^2}{2}\frac{\partial a}{\partial Z}, \qquad (2)$$

$$\frac{\partial \hat{p}_\parallel}{\partial Z} + \frac{g_0^2}{4}\frac{\partial \hat{p}_\parallel}{\partial \tau} = -g_0^2 \frac{\beta_{\perp 0}^2}{2} \text{Re}\left(\frac{\hat{p}_\perp^*}{\hat{p}_\parallel}\frac{\partial a}{\partial Z}\right).$$

Here, the following variables and parameters are used:

$$\tau = \frac{\omega_c \beta_{\perp 0}^4 t}{8\beta_{\parallel 0}^2}, \quad Z = \frac{\beta_{\perp 0}^2 \omega_c z}{2\beta_{\parallel 0} c}, \quad a = \frac{eA J_{m-1}(\kappa R_0)}{mc\omega_c \beta_{\perp 0}^3},$$

$$\hat{p}_\perp = \frac{(p_x + ip_y) e^{-i\omega_c t + i(m-1)\varphi}}{mV_{\perp 0}}, \quad \hat{p}_\parallel = \frac{p_\parallel}{mV_{\parallel 0}}, \quad J = \frac{1}{2\pi}\int_0^{2\pi} \frac{\hat{p}_\perp}{\hat{p}_\parallel} d\theta_0$$

is the amplitude of the RF current, $\beta_{\perp 0} = V_{\perp 0}/c$ and $\beta_{\parallel 0} = V_{\parallel 0}/c$ are the initial transverse and longitudinal velocities of electrons normalized to the speed of light, $g_0 = \beta_{\perp 0}/\beta_{\parallel 0}$ is the initial pitch-factor, $I_0 = 16(eI_b/mc^3)\beta_{\perp 0}^{-6}\beta_{\parallel 0} G$ is the normalized current parameter, $G = J_{m-1}^2(\kappa R_0)(v_n^2 - m^2)^{-1} J_m^{-2}(v_n)$ is the form-factor written for a tubular electron beam with the injection radius $R_0$, $I_b$ is the electron current, $v_n$ is the $n$-th root of the equation $J_m'(v) = 0$, $\Delta = 2(\omega_c - \omega_H)/\omega_c \beta_{\perp 0}^2$ is the initial mismatch between the wave cutoff frequency and the electron gyrofrequency $\omega_H = eH_0/mc\gamma_0$. The term $(g_0^2/4)\partial \hat{p}_\perp/\partial \tau$ in the motion equations permits one to describe an inclination of the beam line with respect to the dispersion characteristic of the waveguide mode [18]. Actually, in the absence of the coupling with the wave, presenting the amplitude the transverse momentum as $\hat{p} \sim \exp(i\Omega\tau - i\Gamma Z)$ (where $\Omega = 8\beta_{\parallel 0}^2(\omega - \omega_c)/\beta_{\perp 0}^4 \omega_c$ is the normalized radiation frequency shift from the cutoff, $\Gamma = 2\beta_{\parallel 0} ck_\parallel/\beta_{\perp 0}^2 \omega_c$ is the normalized longitudinal wavenumber) we obtain for the beam line $\Omega = 4g_0^{-2}(\Gamma - \Delta)$ (see Fig.1).

We assume that at the input cross section $Z = 0$ the electrons are uniformly distributed over cyclotron rotation phases $\hat{p}_\perp(Z = 0) = e^{i\theta_0}$, $\theta_0 \in [0, 2\pi)$ and have the same longitudinal momenta $\hat{p}_\parallel(Z = 0) = 1$. The boundary condition for the wave in the input cutoff narrowing is $a(Z = 0) = 0$. At the system output $Z = L$ (where $L = \beta_{\perp 0}^2 \omega_c l / 2\beta_{\parallel 0} c$ is the normalized length of the interaction space) we apply the well-known reflectionless boundary condition [13].

The electromagnetic wave power near the cutoff frequency is given by the relation $P = (m^2 c^5 / e^2) \beta_{\perp 0}^8 \beta_{\parallel 0}^{-1} G^{-1} \hat{P}$, where $\hat{P} = \mathrm{Im}(a \partial a^* / \partial Z)$ is the normalized energy flux. For description of the pulse generation efficiency, we introduce a conversion coefficient defined (see Ref.19) as the ratio of the radiation power $P$ to the electron beam power $P_{beam} = I_b m c^2 (\gamma_0 - 1)/e$:

$$K = \frac{P}{P_{beam}} = \frac{g_0^2}{1 + g_0^2} \frac{2}{I_0} \hat{P}. \qquad (3)$$

Based on Eqs.(2), simulations of the gyrotron dynamics were performed assuming that the normalized interaction length $L$ is equal to 15, the electron pitch-factor $g_0$ is of 1.3, and $\beta_{\perp 0} = 0.2$ (the electron energy is of 20 keV). As was mentioned above, as the operating current increases, gyrotrons exhibit a transition from steady-state generation with constant amplitude to periodic self-modulation and then to the irregular turbulent state. According to [13,17], the lowest bifurcation values of the current parameter occur for the negative cyclotron resonance detuning $\Delta$. For the case $\Delta = -0.7$ considered below, the generation threshold corresponds to the current parameter $I_0 = 0.004$, a transition to periodic self-modulation takes place for $I_0 = 0.016$, and a chaotic regime occurs for $I_0 \geq 0.1$.

In Fig. 2 we show the output power time traces and histograms of the pulse-height distributions for three values of $I_0$ corresponding to the zone of stochastic generation. One can see that near the boundary of this zone ($I_0 = 0.1$) the ratio of the noise radiation power to its average level $\langle P \rangle$ does not exceed $6 \div 7$ (Fig. 2a). At the same time, as the current parameter increases, isolated giant spikes with peak power $P_{peak}$ much greater than $\langle P \rangle$ appear sporadically in the output signal. For $I_0 = 1.0$, the ratio $P_{peak}/\langle P \rangle$ amounts to $50 \div 100$ (Fig. 2b), while for $I_0 = 3.0$ it reaches $150 \div 200$ (Fig. 2c). The corresponding pulse-height histograms (Fig. 2b,c) have a long-tail profile, with the extreme events occurring frequently than in the case of the relatively narrow distribution for typical events (cf. Fig. 2a). Thus, the pulses generated in gyrotrons exhibit statistical behavior similar to the rogue waves in hydrodynamics and optics [1-9].

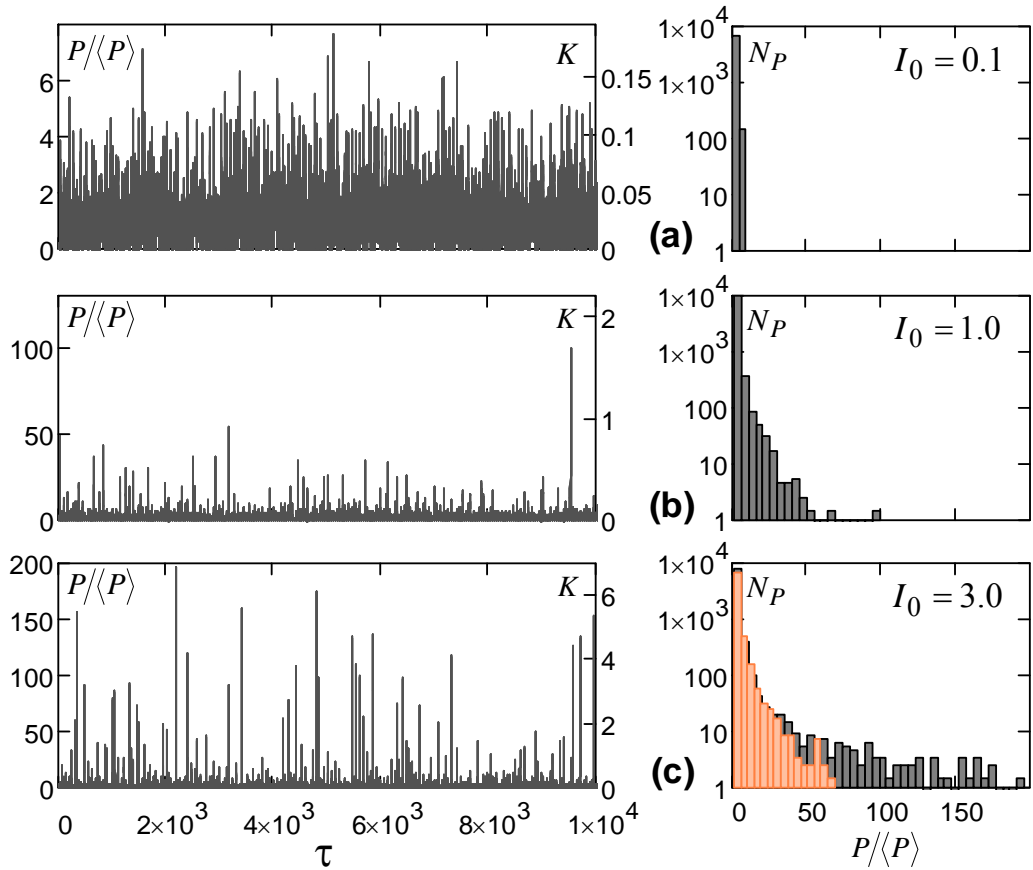

**Fig.2** The output power time traces and histograms of the pulse-height distributions for different values of the normalized current parameter $I_0$ corresponding to the zone of chaotic generation. $N_p$ is the number of spikes in the intensity bins with a given ratio $P/\langle P \rangle$. The distribution shown by grey color in Fig.2c is obtained when the change of a longitudinal momentum is neglected.

The mechanism of the formation of rogue waves in gyrotrons is illustrated by Fig. 3. It is related to a specific operation of gyrotrons in which electrons can interact synchronously with both backward and forward propagating waves near a cutoff frequency. Under a strong excess over the threshold the complex interplay of these mechanisms results in the appearance of giant output pulses. At the preliminary stage, electromagnetic radiation is associated with the excitation of a backward wave (Fig. 3a-c) having a fairly narrow spectrum $S_\Omega = (1/2\pi) \int_{-\infty}^{+\infty} a(\tau) \exp(-i\Omega\tau) d\tau$, corresponding to the low resonance point $I$ in Fig. 1. Upon reflection from the cathode neck, this radiation is partially absorbed by the electron beam, that leads to a dramatic increase in the transverse energy of electrons (Fig. 3e). Thus, we can say that the electrons leaving the interaction space at an early moment of time impart their transverse energy with the backward wave to the later injected electrons which produce a forward propagating pulse. Moving through electrons with large orbital momentum ($g \sim 3.5 \div 4$), such a pulse is effectively amplified while its duration is shortening (cf. [20, 21]). This process is accompanied by the pulse front steepening and excitation of a strong transverse magnetic field $H_\perp \sim \partial a/\partial Z$ (see Fig. 3d). In the area between the cross sections $Z \approx 5$ and $Z \approx 10$ (Fig. 3f) the magnetic field leads to the transformation of the longitudinal to transverse momenta of particles. This provides an additional source of rotational energy for amplification of the forward propagating ultrashort electromagnetic pulse. The appearance of such pulses leads to a significant widening of the output radiation spectrum, which, in fact, is determined by the frequency difference at the resonance points *I* and *II* (Fig. 1, $Z = 15$). It should be noted that a relatively limited part of the electron energy is converted into the energy of electromagnetic radiation. As a result, ejection from the interaction space of electrons with residual transverse momentum significantly exceeding the initial value takes place with a small delay after the emission of a giant electromagnetic pulse Fig. 3e. Nevertheless, the peak power $P_{peak}$ of the formed giant pulses strongly exceeds not only the mean radiation level but, in the optimal case, the kinetic power of the driving electron beam. Pulses with the conversion factor $K \geq 1$ appear in the output radiation for the current parameter $I_0 \geq 1$, and in the range $I_0 \approx 2 \div 4$ this factor reaches the maximum values $K \approx 6 \div 9$.

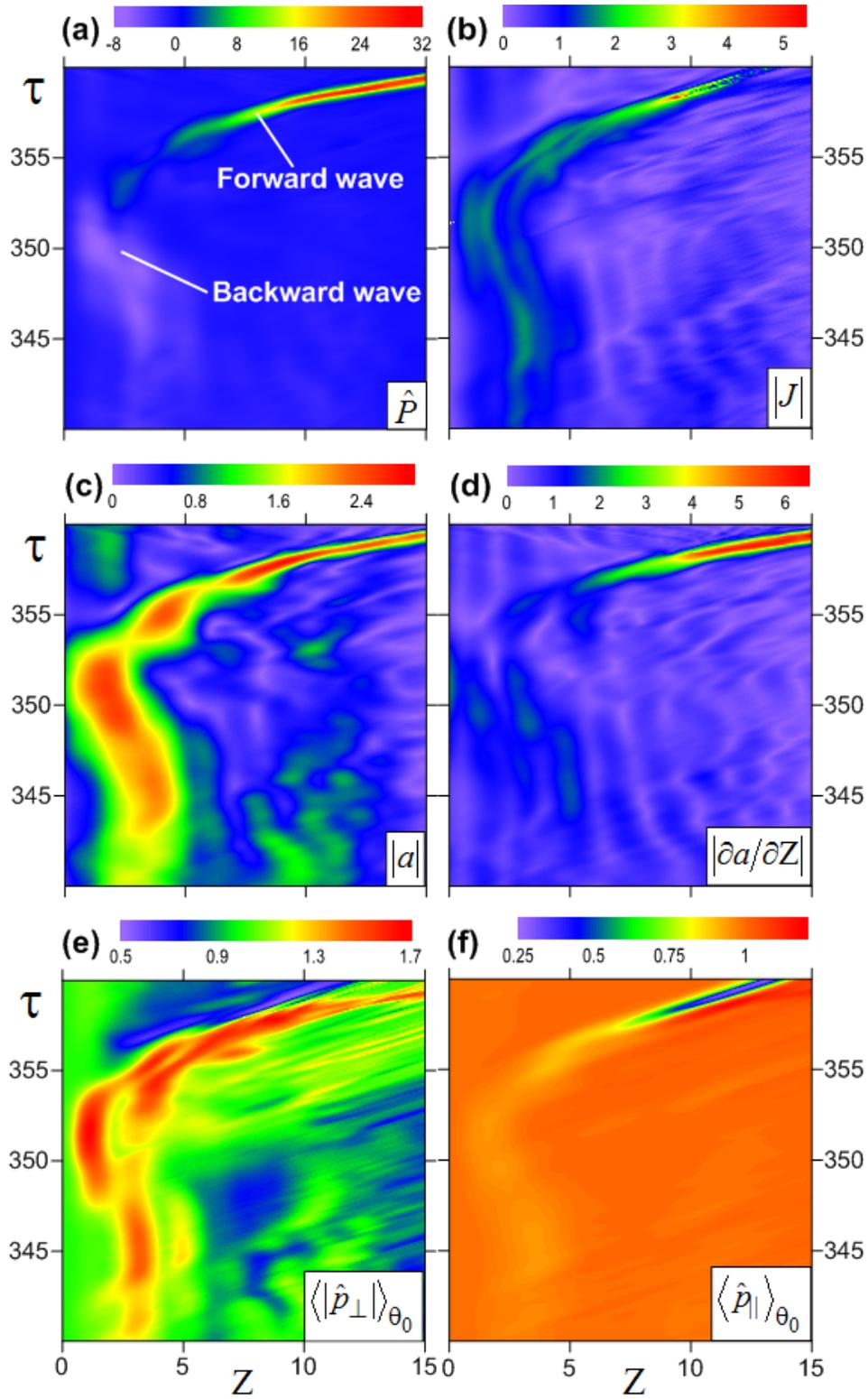

**Fig.3** Evolution of the main physical variables inside the gyrotron interaction space in process of the rogue wave formation: (a) normalized e.m. energy flux $\hat{P} = \mathrm{Im}(a\,\partial a^*/\partial Z)$, (b) the amplitude of the RF current $|J|$, (c) and (d) the amplitudes of the electric $|a|$ and magnetic $|\partial a/\partial Z|$ fields, (e) and (f) the normalized transverse $\langle|\hat{p}_\perp|\rangle_{\theta_0}$ and longitudinal $\langle\hat{p}_\parallel\rangle_{\theta_0}$ momenta of particles averaged on initial cyclotron phases.

The possibility of the rogue wave generation in gyrotrons was confirmed by direct 3D simulations based on the PIC code KARAT [22]. The microwave system of the studied gyrotron is shown in Fig. 4a. Along with the regular section 0.25 cm in radius and 15 cm long, it also includes the cathode narrowing and the collector widening. We consider the excitation of the $TE_{11}$ mode with a cutoff frequency of 35 GHz by a tubular electron beam with parameters typical for millimeter waveband gyrotrons (see, e.g., [23]): particles energy of 20 keV, an electron current of 2 A, and a pitch-factor of 1.3. The guiding homogeneous magnetic field is of 13 kOe. In this case, the corresponding normalized parameters are basically the same as those used in the simulations presented in Fig. 2c. However, a 20% spread of the orbital electron velocities was taken into account in the PIC simulations.

The results of the PIC simulations are presented in Fig. 5 showing the time dependence of the output power, the corresponding long-tail pulse-height distributions, and the radiation spectrum. It is seen that the gyrotron radiation is characterized by appearance of high-amplitude ultrashort pulses with durations of 0.2 to 0.25 ns and intervals of 20 to 100 ns. The peak power of generated pulses reaches $200 \div 230$ kW, which corresponds to the conversion factor of $4.8 \div 5.5$. The ratio of the pulse peak power to the mean radiation power reaches 125, that is in good agreement with the results obtained on the basis of Eqs.(2). It should be noted that formation of the backward wave (Fig. 4b) and strong axial deceleration of electrons (Fig. 4c) was also observed in the PIC simulations.

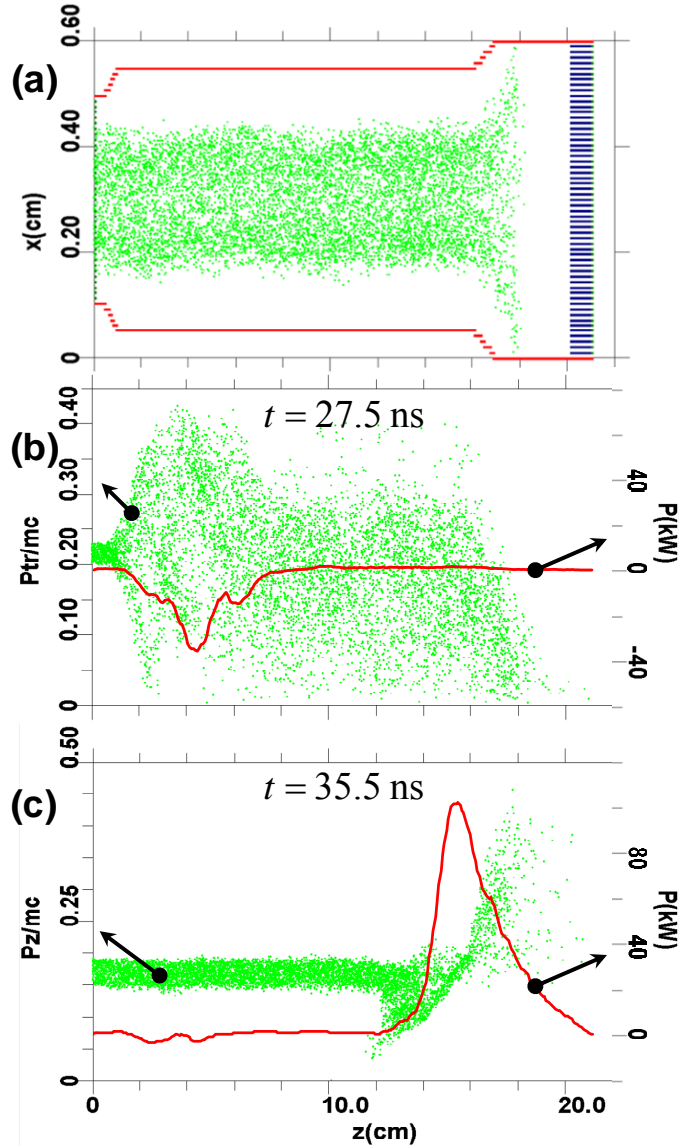

**Fig.4** (a) Geometry of a 35-GHz gyrotron used in the 3D PIC simulations and instantaneous positions of macroparticles ($U = 20$ kV, $I = 2$ A, $g_0 = 1.3$, $H_0 = 13$ kOe). (b) Increase of transverse momenta of electrons in the field of the backward wave. (c) Typical spatial profile of the forward propagating giant spike and longitudinal deceleration of electrons in the interaction space.

Thus, in this paper we demonstrate that in the gyrotrons operating in the regime of developed chaos, high-power ultrashort microwave pulses can be sporadically emitted from the interaction space. We use the term "rogue waves" for these pulses to outline their physical similarities with the extreme events in optical systems and hydrodynamics. Moreover, the rogue waves in gyrotrons can reach intensities at least a factor of $100 \div 150$ greater than the average radiation power, which significantly exceeds typical values for the extreme events in lasers [5-7].

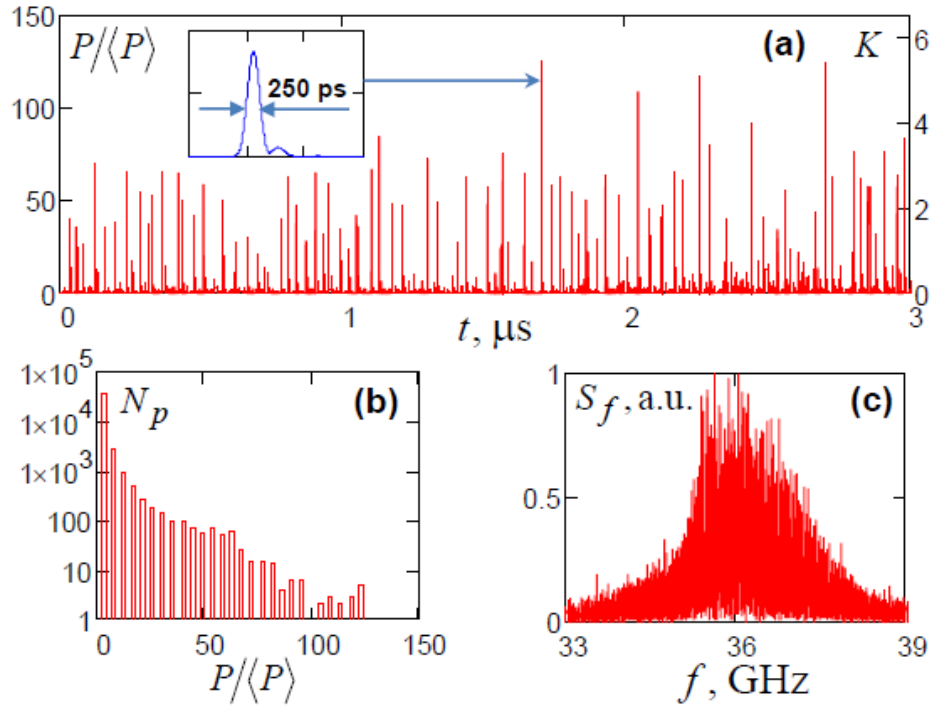

**Fig.5** Results of the PIC simulations: (a) the time dependence of the output power and typical temporal profile of a generated spike; (b) the corresponding long-tail pulse-height distribution, and (c) the broadband radiation spectrum.

From the practical point of view, experimental observation of the rogue waves generation in gyrotrons can be interesting as a method for production of high-power millimeter-wave radiation with an ultrawide spectrum, which, according to our PIC simulations, amounts to 10% at the level of -10 dB (Fig. 5c). It should be noted that the rogue waves generation can be realized with routine parameters of the driving electron beams formed by the magnetron-injection guns widely used in gyrotons. The required exceeding over the threshold can be provided due to operation at the lowest-order modes of the microwave cavities.

The authors are grateful to E.N. Pelinovsky for stimulating discussions. This work was supported by the Russian Science Foundation (project No. 16-42-01078).

**REFERENCES**

[1] Extreme Ocean Waves, Ed. by E. Pelinovsky, C. Kharif. – Springer Science+Business Media B.V., 2008.

[2] M. Onorato, S. Residori, U. Bortolozzo, A. Montina, and F.T. Arecchi, Phys. Reports **528**, 47 (2013).

[3] A.N. Ganshin, V.B. Efimov, G.V. Kolmakov, L.P. Mezhov-Deglin, and P.V.E. Mcclintock, Phys. Rev. Lett. **101**, 065303 (2008).

[4] D.R. Solli, C. Ropers, P. Koonath, and B. Jalali, Nature **450**, 1054 (2007).

[5] C. Lecaplain, Ph. Grelu, J.M. Soto-Crespo, and N. Akhmediev, Phys. Rev. Lett. **108**, 233901 (2012).

[6] F. Selmi, S. Coulibaly, Z. Loghmari, I. Sagnes, G. Beaudoin, M.G. Clerc, and S. Barbay, Phys. Rev. Lett. **116**, 013901 (2016).

[7] J. Peng, N. Tarasov, S. Sugavanam, and D. Churkin, Opt. Express **24**, 24256 (2016).

[8] R. Sabry, W.M. Moslem, and P.K. Shukla, Phys. Rev. E **86**, 036408 (2012)

[9] R. Hohmann, U. Kuhl, H.-J. Stockmann, L. Kaplan, and E.J. Heller, Phys. Rev. Lett. **104**, 093901 (2010).

[10] G.S.Nusinovich, "Introduction to Physics of Gyrotons", J. Hopkins Univ. Press, 2004.

[11] N.S. Ginzburg, S.P. Kuznetsov, and T.N. Fedoseeva, Radiophys. & Quant. Electronics **21**, 728 (1978) (DOI: 10.1007/BF01033055).

[12] N.S. Ginzburg, N.I. Zaitsev, E.V. Ilyakov, I.S. Kulagin, Yu.V. Novozhilova, R.M. Rozenthal, and A.S. Sergeev, Phys. Rev. Lett. **89**, 108304 (2002).

[13] N.S. Ginzburg , G.S. Nusinovich, and N.A. Zavolsky, Int. J. of Electronics **61**, 881 (1986).

[14] K.F. Pao, T.H. Chang, C.T. Fan, S.H. Chen, C.F. Yu, and K.R. Chu, Phys. Rev. Lett. **95**, 185101 (2005).

[15] E.V. Blokhina, S.P. Kuznetsov, and A. G. Rozhnev, IEEE Trans. on Electron Devices **54**, 188 (2007).

[16] S. Alberti, F. Braunmueller, T.M. Tran, J. Genoud, J-Ph. Hogge, M.Q. Tran, and J-Ph. Ansermet, Phys. Rev. Lett. **111**, 205101 (2013).

[17] O. Dumbrajs1 and G.S. Nusinovich, Phys. Plasmas **23**, 083125 (2016).

[18] N.S. Ginzburg, A.S. Sergeev, and I.V. Zotova, Phys. Plasmas **22**, 033101 (2015).

[19] S.D. Korovin, A.A. Eltchaninov, V.V. Rostov, V.G. Shpak, M.I. Yalandin, N.S. Ginzburg, A.S. Sergeev, and I.V.Zotova, Phys.Rev.E **74**, 016501 (2006).

[20] T-B.Zhang, and T.C.Marshall, Phys.Rev.Lett. **74**, 916 (1995).

[21] M.I. Yalandin, A.G. Reutova, M.R. Ul'maskulov, K.A. Sharypov, S.A. Shunailov, N.S. Ginzburg, I.V. Zotova, E.R. Kocharovskaya, and A.S. Sergeev, JETP Letters, **91**, 553 (2010).

[22] V.P. Tarakanov, User's Manual for Code KARAT (Springfield, BRA, 1992).

[23] Yu. Bykov, A. Eremeev, M. Glyavin, V. Kholoptsev, A. Luchinin, I. Plotnikov, G. Denisov, A. Bogdashev, G. Kalynova, V. Semenov, and N. Zharova, IEEE Trans. Plasma. Sci. **32**, 67 (2004).